\documentclass[aps,amsmath,amssymb,twocolumn,showpacs,showkeys,fleqn]{revtex4-1}

\usepackage{graphicx}
\DeclareGraphicsExtensions{.jpg}
\usepackage{dcolumn}
\usepackage{bm}
\usepackage{amssymb}
\usepackage{amsmath}
\usepackage{bm}
\usepackage[nodots]{numcompress}
\usepackage{gensymb}
\usepackage[numbered]{bookmark}
\usepackage{comment}
\usepackage{caption}
\usepackage{subcaption}
\usepackage{float}
\usepackage{setspace}
\usepackage[sort&compress]{natbib}

\begin{document}


\title{Ferrofluid nucleus phase transitions in an external uniform magnetic field}


\author{B. M. Tanygin}
 \email{b.m.tanygin@gmail.com}

\author{S. I. Shulyma}

\author{V. F. Kovalenko}

\author{M. V. Petrychuk}

\affiliation{Faculty of Radiophysics, Taras Shevchenko National University of Kyiv, 4G, Acad. Glushkov Ave., Kyiv, Ukraine, 03187}



\date{\today}

\begin{abstract}

The phase transition between a massive dense phase and a diluted superparamagnetic phase has been studied by means of a direct molecular dynamics simulation. The equilibrium structures of the ferrofluid aggregate nucleus are obtained for different values of a temperature and an external magnetic field magnitude. An approximate match of experiment and simulation has been shown for the ferrofluid phase diagram coordinates ``field-temperature''. The provided phase coexistence curve has an opposite trend comparing to some of known theoretical results. This contradiction has been discussed. For given experimental parameters, it has been concluded that the present results describe more precisely the transition from linear chains to a dense globes phase. The theoretical concepts which provide the opposite binodal curve dependency trend match other experimental conditions: a diluted ferrofluid, a high particle coating rate, a high temperature, and/or a less particles coupling constant value.
\end{abstract}


\maketitle


\section{Introduction}
\label{ref_intro}


Ferrofluids (FF) are colloidal suspensions of magnetic nanoparticles in a carrier liquid. An increasing interest to FF is related to their applications: drug deliveries \cite{Lubbe1996}, a hyperthermia \cite{Jordan1999a}, a magneto-resonance tomography \cite{Kim2001}, others \cite{Shimada2005,Yellen2004,Li2014,YuGuo-Jun}. Additionally, phase transitions fundamental understanding can be developed by means of FF usage as a model system.

Suspended and aggregated phases are two major possible states of a FF matter \cite{Blum1986,HOLM2005}. A vessel volume contains a single phase or multiple FF phases \cite{Zubarev2004b}. A more detailed systematization of dense phases is determined by an aggregates substructure classification (spatial ordering and symmetry) which impacts on thermodynamic properties. Some of possible FF aggregates are: drop-like aggregates \cite{Bacri1982,Zubarev2003,Zubarev2004b}(either bulk drops or microdrops), one-dimensional chains \cite{Iskakova2006,Tanygin2012e,Yoon2010,GORENSTEIN2014}, labyrinthine patterns \cite{rosensweig1983labyrinthine}, hexagonal patterns \cite{Richardi2004}, rod-shaped \cite{Moldovan2006a,Petrychuk2006} aggregates, dumbbell-like aggregates \cite{Tanygin2013,Petrychuk2006}, others. Phase transitions correspond to all possible pairs of these phases. Forced phase transitions can be triggered by an external magnetic field applying and/or temperature changes or through other external impact.

The phase transition between the suspended and aggregated phases has been called by a liquid-gas (l-g) phase transition \cite{HOLM2005}. A contradiction between different FF l-g phase transition research results exists \cite{HOLM2005}: while it is predicted by theories \cite{sano1983theory,YuZubarev2005,Ivanov1996,Li2007,Zubarev2004b,Kalikmanov1992} and observed in some experiments \cite{Cousin2003}, simulation research reports (e.g. \cite{Weis1993}) usually exclude the l-g phase transition possibility or provides a phase diagram which does not match an expected one \cite{HOLM2005}. It has been concluded that mean-field and statistical models are not justified in the case of large coupling constant and/or large densities of FF nanoparticles \cite{Wang2002}. Other aspects of the above-mentioned contradiction will be discussed in the present work.

A material memory related phenomenon is detected experimentally for drop-like aggregates. The phase diagram evolves after a cyclic heating and cooling: temperature and field magnitude transition values depend on a sequential cycle number \cite{Kovalenko2013,Kovalenko2014}. This effect as well as similar ones cannot be explained in scope of theories which imply a FF as a continuous medium of separate particles. Such thermodynamical (statistical) theories \cite{tsebers1982role,sano1983theory,Scholten1983,Cebers1990,Buyevich1992,Kalikmanov1992,Ivanov1996,Li2007} require assumptions leading to an analytical closed-form expression derivation possibility. Some of these theories are phenomenological (e.g. \cite{Li2007}). Examples of the simplification related assumptions of these models are following: the nearest-neighbors approximation of configurational integral \cite{Zubarev2004b}, the isotropic potential approximation \cite{Chan1984}, a dipole-dipole interaction as a perturbation \cite{Ivanov1996}, a very diluted phase consideration \cite{Stevens1994,Satoh1996}, a zero or infinite value of an external magnetic field \cite{Zubarev2004b}, others \cite{Mansoori1971}. Some of these models \cite{Chan1984} imply the Boltzmann factor based statistical averaging over all possible magnetic moments pair orientations. However, this model assumes an uniform statistical distribution over particles mutual positions which is justified only in the case of a high-temperature FF with a small correlation of magnetic moments orientations \cite{Kalikmanov1992}. This assumption works in the case of a temperature close or higher than a critical temperature of a FF. Similar model shows matching an experiment in the case of a diluted enough FF: a volume fraction value of a FF dispersed phase was $2.4\,\%$ \cite{Ivanov1996}. This match has been shown only for a sedimentation direction of a phase transition in contrast of a ``melting'' \cite{Scholten1983} of a dense phase.

A short review and comparison of phase transitions investigations in theoretical, experimental and numerical simulation research works has been reported in the~\cite{Zubarev2004b}. The idealized ``homogeneous'' FF models treats the phase transition as the van der Waals l-g transition of an ensemble of separate particles. This assumption leads to the problems of matching of the theory and the experiment \cite{Zubarev2004b}. Any heterogeneous clusters (e.g. one-dimensional chains or coil-like structures) are not covered by such models. A detailed study of FF aggregates equilibrium internal structure shows a much wider variety of possible structures \cite{Yoon2010,Tanygin2012e}. A thermodynamic stability of FF aggregate structures leads to an emergence of a gap between branches of a binodal of a coexistence of dense and dilute phases \cite{Zubarev2004b}.

In order to take into account a variety of possible spatial and spin configurations of a dipolar system, a numerical modeling based research work is required. There are two major types of such methods: molecular dynamics (see e.g. \cite{Wang2002,Wang2003} and references therein) and Monte Carlo (see e.g. \cite{Satoh1996,Camp2000} and references therein) simulations. In scope of a conventional assumption that the Brownian particles motion does not require a quantum mechanics based description \cite{Hakim1985}, a molecular dynamics is a possible candidate to be an \textit{ab initio} method of a FF numerical research. Another assumption is implicit solvation (continuum solvation) \cite{Ferrara2002} approximation of Brownian particle motion. Most molecular dynamics simulations focus on systems with periodical boundary conditions \cite{Wang2002,Wang2003}. A comprehensive comparison of a theory, a simulation, and an experiment has been reported for magnetization curves only \cite{Ivanov2007}. Concerning phase diagrams, such comparison is either non-quantitative or requires an introduction and a variation of \textit{ad hoc} unknown parameters leading to an expected match: magnetic cores size distribution functions (a polydispersity), a nonspherical particle shape, the Hamaker constant $A_\mathrm{H}$ precise value definition problem \cite{Scholten1983}, the van der Waals forces microscopic theory approximation including many-body interactions \cite{Derjaguin1934a,Lifshitz1956,Langbein1970a}, a nonmagnetic part of a nanoparticle volume, a stabilizing surfactant layer interaction nature (formula and constants) \cite{Wang2002}, a coverage rate of a nanoparticle by a surfactant, solvation layers \cite{Ivanov1996}, aggregating characteristic parameter \cite{Li2007}, etc. There is a lack of numerical research works which match experimental phase transitions parameters \cite{Ke2004,Morozov1987,Bacri1989,Bacri1990,Kovalenko2013,Kovalenko2014} even via the selection of the above-mentioned \textit{ad hoc} parameters. According to~\cite{Zubarev2004b}, existing simulation methods \cite{Wang2002,Wang2003,Polyakov2013,caillol1993search,van1993makes,levesque1994orientational,Stevens1994,Satoh1996} provides mostly linear chain-like aggregates. Hence, the phase transition between a massive dense phase \cite{Tanygin2012e} and a diluted superparamagnetic phase is not yet enough studied by a direct molecular dynamics simulation.

Our previous published simulation method had been designed to the purpose of a monodisperse FF aggregation research in the case of large magnetite particles (diameter $20$ nm) \cite{Tanygin2012e}. A smaller size of magnetite particles ($\approx 10$ nm or less) corresponds to a predominance of the Brownian motion comparing to aggregating forces \cite{Fertman1990,shliomis1974magnetic}. Hence, in real polydisperse FF complex phase transitions and phases coexistence are possible \cite{Zubarev2004b,Ivanov1996}. The \textit{subject} of the present research is the phase transitions in the polydisperse FF under applied temperature and magnetic field changes. For this purpose, the original simulation method should be improved and validated by a comparison with related experimental research \cite{Kovalenko2013,Kovalenko2014}.

\section{Model and simulation}
\label{ref_sim}

\subsection{Particle model and interactions}

If a deviation of a nanoparticle shape from a spherical one has isotropic probability distribution then one could consider particles as hard spheres \cite{Fertman1990}. The $i$-th particle hydrodynamical diameter $d_i = d_i^{\mathrm{m}} + 2 a_0 + 2 \delta$ consists of a ferromagnetic core diameter $d_i^{\mathrm{m}}$, a nonmagnetic surface layer thickness which equals a lattice constant $a_0$ of a respective bulk crystalline material, and a stabilizing surfactant layer thickness $\delta$ \cite{Fertman1990,shliomis1974magnetic}. We will consider only single-domain particles with $d_i^{\mathrm{m}} < d_0$ \cite{Brown1968} where the single-domain particle size threshold is $d_0 < 0.5\mu$m \cite{Dunlop1972,Dunlop1973,Butler1975}. A stabilizing surfactant molecules surface density $N_{\mathrm{S}}$ on a particle determines a rate of its coating and impacts on the FF stabilization \cite{Fertman1990}.

Lagrangian and Rayleigh dissipation function of a $N$ particles system include each particle degrees of freedom ($i=\overline{1,N}$): center of mass spatial coordinates $\bm{r}_i$, Euler angles of a particle rotation, and a magnetic moment $\bm{m}_i$ where $\left|\bm{m}_i\right|=const$. A particle potential energy depends on its Euler angles implicitly through an angle $\theta_i$ between $\bm{m}_i$ and a particle easy magnetization axis unit vector $\bm{n}_i$ \cite{shliomis1974magnetic}.

The total force acting on the $i$-th nanoparticle is given by:
\begin{equation} \label{eq:total_f}
	\bm{F}_i = \bm{F}_i^{\mathrm{f}} + \sum_{j}\left( \bm{F}_{ij}^{\mathrm{dd}} + \bm{F}_{ij}^{\mathrm{W}} + \bm{F}_{ij}^{\mathrm{ER}} + \bm{F}_{ij}^{\mathrm{HS}} \right)
\end{equation}
Assuming the limit of low Reynolds number, the viscous friction force is given by the Stokes' law $\bm{F}_i^{\mathrm{f}} = - 3 \pi \eta d_i \bm{v}_i \equiv - \gamma^{\mathrm{T}}_i \bm{v}_i$ where the $\eta$ is a carrier liquid dynamic viscosity and the $\bm{v}_i$ is the $i$-th particle velocity. The $\bm{F}_{ij}^{\mathrm{dd}}$ is a dipole-dipole interaction force \cite{shliomis1974magnetic,Tanygin2012e,Wang2002}. The van der Waals' interaction energy between spherical bodies is given in \cite{hamaker1937london}. Its differentiation yields the force:
\begin{equation} \label{eq:van_der_W}
	\bm{F}_{ij}^{\mathrm{W}} = \boldsymbol{\rho}_{ij} \frac{-32 \,A_{\mathrm{H}}\,z\,{R_i}^{3}\,{R_j}^{3}}{3\,{\left[ {z}^{2}-{\left( R_i-R_j\right) }^{2}\right] }^{2}\,{\left[ {z}^{2}-{\left( R_j+R_i\right) }^{2}\right] }^{2}}
\end{equation} where $A_{\mathrm{H}}$ is the Hamaker constant; a radius $R_i \equiv a_0 + d_i^{\mathrm{m}} / 2$ ; a distance $z = \left|\bm{r}_i - \bm{r}_j\right|$; and its dimensionless value $\boldsymbol{\rho}_{ij} = \left(\bm{r}_i - \bm{r}_j\right) / z$.
The energy density of entropic repulsion \cite{rosensweig1997ferrohydrodynamics} of two surfaces:
\begin{equation} \label{eq:entropic_rep_sf}
E_{\mathrm{f}} =k\,T\,N_{\mathrm{S}}\,\left[ 1-s / (2\,\delta)\right]
\end{equation}
after an integration over surfaces of two particles with different radii takes the form (at $z \leq R_i + R_j + 2 \delta$):
\begin{widetext}
\begin{equation} \label{eq:entropic_rep}
	G_{ij}^{\mathrm{ER}} = \frac{\pi k T N_{\mathrm{S}}{\left[ z-\left( 2\,\delta+{R}_{j}+{R}_{i}\right) \right] }^{2}\,\left[ \left( {R}_{j}+{R}_{i}\right)^2 \,\left( z+\delta\right) -({R}_{i}^{3}+{R}_{j}^{3})\right]}{6\,\delta\,z\,({R}_{i}+{R}_{j})}
\end{equation}
where $k$ is the Boltzmann constant; $T$ is a thermodynamic temperature; $s$ is a distance between surfaces; $\delta$ is a surfactant molecule length. In the case of particles of same radii, the closed-form expression equation (\ref{eq:entropic_rep}) differs from the well-known logarithmic one~\cite{rosensweig1997ferrohydrodynamics,rosensweig1965ferrohydrodynamic}. However, these dependencies are numerically close to each other with the tolerance $\sim 20~\%$ for particles with diameter $10$~nm. Existing experimental results cannot justify these expressions difference. The comparison of the derivation procedures will be published separately.

The corresponding force $\bm{F}_{ij}^{\mathrm{ER}} = - \partial G_{ij}^{\mathrm{ER}} / \partial \bm{r}_i$ is given by:
\begin{align}
\bm{F}_{ij}^{\mathrm{ER}} &= -\boldsymbol{\rho}_{ij} \pi k T N_{\mathrm{S}}\left( z-2\delta-{R}_{2}-{R}_{1}\right)\left( {R}_{2}+{R}_{1}\right) \times \nonumber\\
&\times \left\{  2\left( {z}^{2}+{\delta}^{2}\right) -\left( {R}_{2}^{2}-{R}_{1}{R}_{2}+{R}_{1}^{2}\right) \left( \frac{z}{{R}_{2}+{R}_{1}}+1\right) +\delta z
+ \frac{\left[ 2{R}_{1}{R}_{2}-{\left( {R}_{1}-{R}_{2}\right) }^{2}\right] \delta}{{R}_{2}+{R}_{1}}\right\} 
 / \left(6\delta{z}^{2}\right)
\end{align}
The hard-sphere model \cite{Kalikmanov1992} force is given by the following expression:
\begin{equation}
	\bm{F}_{ij}^{\mathrm{HS}} =  \left\{
																	 \begin{array}{c}
																	 \infty \cdot \boldsymbol{\rho}_{ij},\ \mathrm{at}\ z \leq (d_i + d_j) / 2 \\
																	 \bm{0},\ \ \mathrm{at}\ z > (d_i+d_j)/2 \\
																	 \end{array}
														\right.
\end{equation}
The total effective torque acting on the $i$-th particle is determined by a viscous rotational friction, a particle anisotropy, the applied external magnetic field $\bm{B}_0$ value and a dipole-dipole interaction field \cite{shliomis1974magnetic,Wang2002,Polyakov2013}:
\begin{equation} \label{eq:total_tau}
	\boldsymbol\tau_i = - \gamma^{\mathrm{R}}_i \boldsymbol\omega_i + K\, V_i\, \mathrm{sin}\left(2 \theta_i \right)\, \left[\bm{n}_i \times \bm{m}_i\right] / \mathrm{sin}\left(\theta_i\right) + \left[\bm{m}_i \times \left(\bm{B}_0 + \sum_{j} \bm{B}_j \right)\right]
\end{equation}
where the $\gamma^{\mathrm{R}}_i = \pi\, d_i^3\, \eta\,$; $V_i = \pi (d_i^{\mathrm{m}})^3 / 6$ is a $i$-th particle magnetic core volume; $K$ is a particle magnetic anisotropy constant; the $\boldsymbol\omega_i$ is the $i$-th particle self rotation angular velocity; the $\bm{B}_j$ is a magnetic field created by the $j$-th particle magnetic moment in the geometric center of $i$-th particle. All types of magnetic anisotropies should be taken into account: a magnetocrystalline anisotropy \cite{brown1963micromagnetics}, a shape anisotropy (demagnetization tensor) \cite{Tandon2004}, an anisotropy of surface \cite{Aharoni1987}, etc.
\end{widetext}
\subsection{Translational and rotational motion}

In scope of the continuum solvation approximation of not very dense solutions, the Brownian translational and rotational particle motion is described by the Langevin equations with the hydrodynamic-originated Langevin parameters \cite{schlick2010molecular,Pottier2014,Wang2002,Wang2003,Polyakov2013}:

\begin{align}
	&M_i \mathrm{d}^2\bm{r}_i/\mathrm{d}t^2 = \bm{F}_i + \boldsymbol\xi_i^{\mathrm{T}} \label{eq:t_motion} \\
	&I_i \mathrm{d}\boldsymbol\omega_i/\mathrm{d}t = \boldsymbol\tau_i + \boldsymbol\xi_i^{\mathrm{R}} \label{eq:r_motion}
\end{align}
where the $M_i$ and $I_i$ are the $i$-th particle mass and moment of inertia respectively; the $t$ is a time; the $\boldsymbol\xi_i^{\mathrm{T}}$ and $\boldsymbol\xi_i^{\mathrm{R}}$ are a random force and torque respectively, which are usually modelized by Gaussian noise \cite{Wang2002,Wang2003,Pottier2014,Mazur1991}:
\begin{align}
	&\left\langle \boldsymbol\xi_i^{\mathrm{T}}\left(t\right)\right\rangle = 0 \label{eq:rnd_force_1}\\
	&\left\langle \boldsymbol\xi_i^{\mathrm{T}}\left(t\right)\boldsymbol\xi_i^{\mathrm{T}}\left(t'\right)\right\rangle = 6 k_{\mathrm{B}} T \gamma^{\mathrm{T}}_i \delta\left(t-t'\right) \label{eq:rnd_force_2}\\
	&\left\langle \boldsymbol\xi_i^{\mathrm{R}}\left(t\right)\right\rangle = 0 \label{eq:rnd_force_3}\\
	&\left\langle \boldsymbol\xi_i^{\mathrm{R}}\left(t\right)\boldsymbol\xi_i^{\mathrm{R}}\left(t'\right)\right\rangle = 6 k_{\mathrm{B}} T \gamma^{\mathrm{R}}_i \delta\left(t-t'\right) \label{eq:rnd_force_4}
\end{align} where $\delta\left(t\right)$ is the Dirac delta.

It is important to note that original Langevin equations were supplemented by particles interaction forces~(\ref{eq:total_f}) and torques~(\ref{eq:total_tau}). This supplementation has been considered as an obvious and intuitive step which had been made in scope of the molecular dynamics simulation based research programs \cite{Wang2002,Wang2003,Polyakov2013}. However, it still should be exactly theoretically justified in general case of the Brownian particle motion. This statement is in need of further research.

A rotation of a magnetic moment inside a particle is described by the Landau-Lifshitz-Gilbert equation with the attempt time $t_0=M_{\mathrm{s}}/{2 \alpha \gamma K}$ where $M_{\mathrm{s}}$ is a saturation magnetization; the $\alpha$ is a damping factor; the $\gamma$ is an electron gyromagnetic ratio \cite{shliomis1974magnetic}. The simultaneous Brownian dynamics of a magnetic moment and a particle has been investigated in the~\cite{Cebers1975}.

\subsection{Experimental and simulation parameters}

All parameters of the present simulation correspond to the specific experiment \cite{Kovalenko2013,Kovalenko2014,Padalka2004}. The considered FF consists of the magnetite nanoparticles suspended in the kerosene carrier liquid. The stabilizing surfactant is the oleic acid ($\delta = 2$ nm~\cite{Fertman1990}). The particles diameter distribution was determined experimentally \cite{Padalka2004}. The mean diameter is $\bar{d} = 11.5$ nm. The magnetite lattice parameter $a_0 \approx 0.8397$ nm corresponds to the cubic spinel structure with the space group Fd3m (above the Verwey temperature) \cite{Fleet1981,el2012synthesis}. However, this parameter is approximate because only 10\% of particles have a crystal structure. This conclusion was made based on a dark field electron microscopy measurements \cite{Padalka2004}. The simulation initial condition is a random close packing \cite{Song2008} (cf. \cite{Scholten1983}) of particles positions and random magnetic moments directions. The external magnetic field is uniform. Dielectric properties of the surfactant layer are generally similar to those of the carrier liquid~\cite{Scholten1983}. Hence, the Hamaker constant of a magnetite is $A_{\mathrm{H}} \, = \, 4 \cdot 10^{-20}$ J \cite{Scholten1983}.

A required step of a dense phase emergence is an original phase nucleus forming. Consequently, a phase diagram of a nucleus is close to a phase diagram of a bulk FF phase. Only the nucleus aggregate will be considered in this simulation. Hence, a number of particles $N$ should be minimal but not less some threshold where a phase diagram start significant changes depending on the $N$: a transition from a bulk phenomenon to a surface one. Number $N \sim 100$ has been selected for the polydisperse FF with a lognormal distribution.

The magnetic moment precession attempt (damping) time order of magnitude is $\tau_0 \sim 10^{-10}-10^{-9}$ s \cite{Fertman1990}. In the case of particles $d_i \leq \bar{d}$, the N\'eel relaxation time $\tau_{\mathrm{N}}$ and the Brownian relaxation time $\tau_{\mathrm{B}}$ relates as \cite{Fertman1990,rosensweig1997ferrohydrodynamics}:
\begin{equation} \label{eq:neel_particles}
	\tau_{\mathrm{N}} < \tau_{\mathrm{B}} \leq ~10^{-6} (\mathrm{s})
\end{equation}
Most part of the range $0 < d \leq \bar{d}$ corresponds to the relation $\tau_{\mathrm{N}} << \tau_0$. In this case both the N\'eel relaxation flip and a magnetic moment dynamics should be considered. The Brownian rotation is a much slower process. A particle rotation does not impact on a magnetic configuration and a free energy of the phase.

The relation is opposite for the larger particles $d_i > \bar{d}$ \cite{shliomis1974magnetic}:
\begin{equation} \label{eq:brown_particles}
	\tau_0 \ll ~10^{-6} (\mathrm{s}) < \tau_{\mathrm{B}} < \tau_{\mathrm{N}}
\end{equation}
Here, a magnetic moment alignment with an effective field can be modeled as instant. The magnetic moment rotates the particle through anisotropy forces. We suppose that the high enough anisotropy energy $K V$ gradient leads to the model with the magnetic moment ``frozen'' into the particle. The vector $\bm{m}_i$ is followed by the vector $\bm{n}_i$:
\begin{equation} \label{eq:frozen}
	\bm{m}_i\,||\,\bm{n}_i
\end{equation}

\begin{figure}[H]
	\centering
	\begin{subfigure}[t]{0.5\textwidth}
		\centering
		\includegraphics[width=\textwidth]{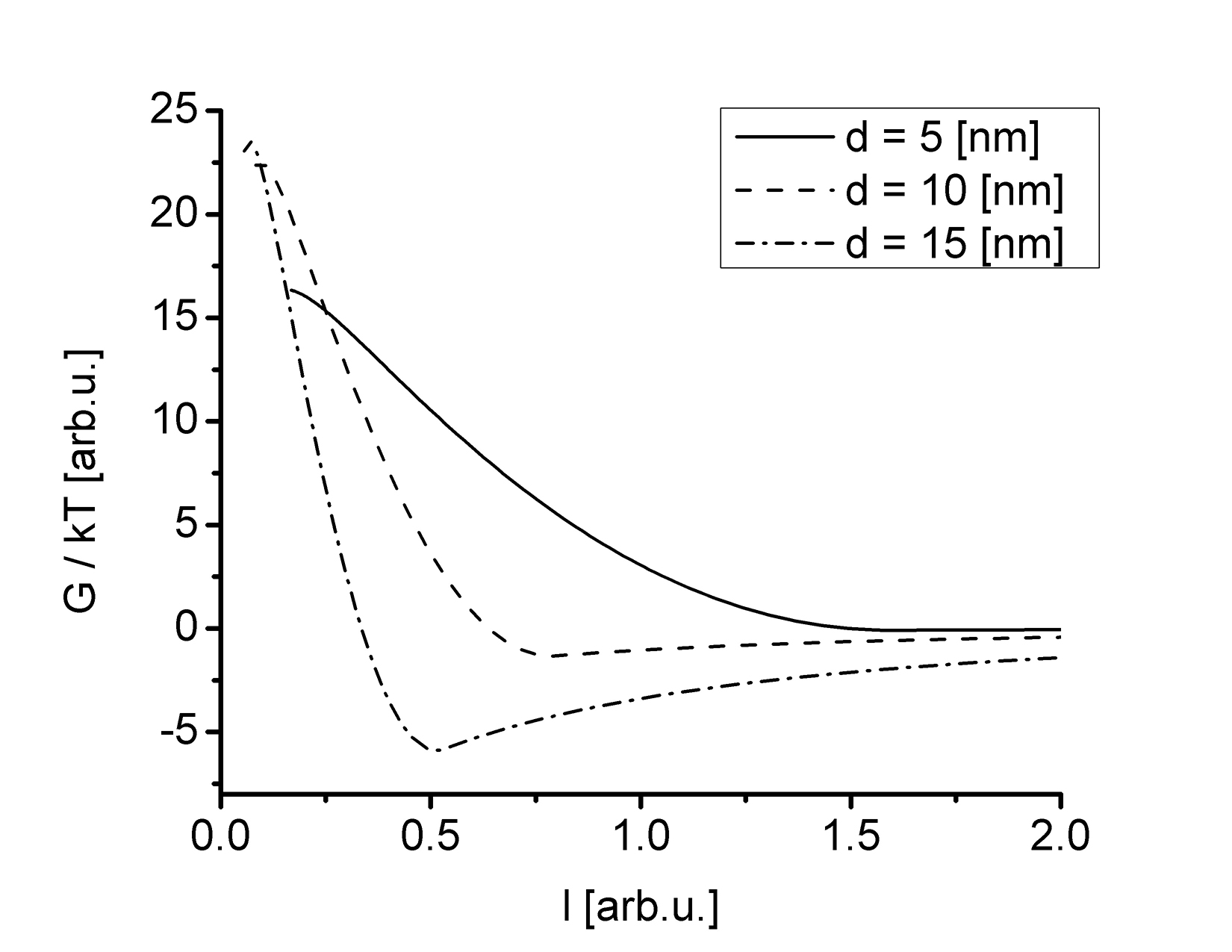}
		\caption{Coating rate by the oleic acid: $k_{\mathrm{c}} = 50\%$}\label{fig:pot_energy_50}		
	\end{subfigure}
	\quad
	\begin{subfigure}[t]{0.5\textwidth}
		\centering
		\includegraphics[width=\textwidth]{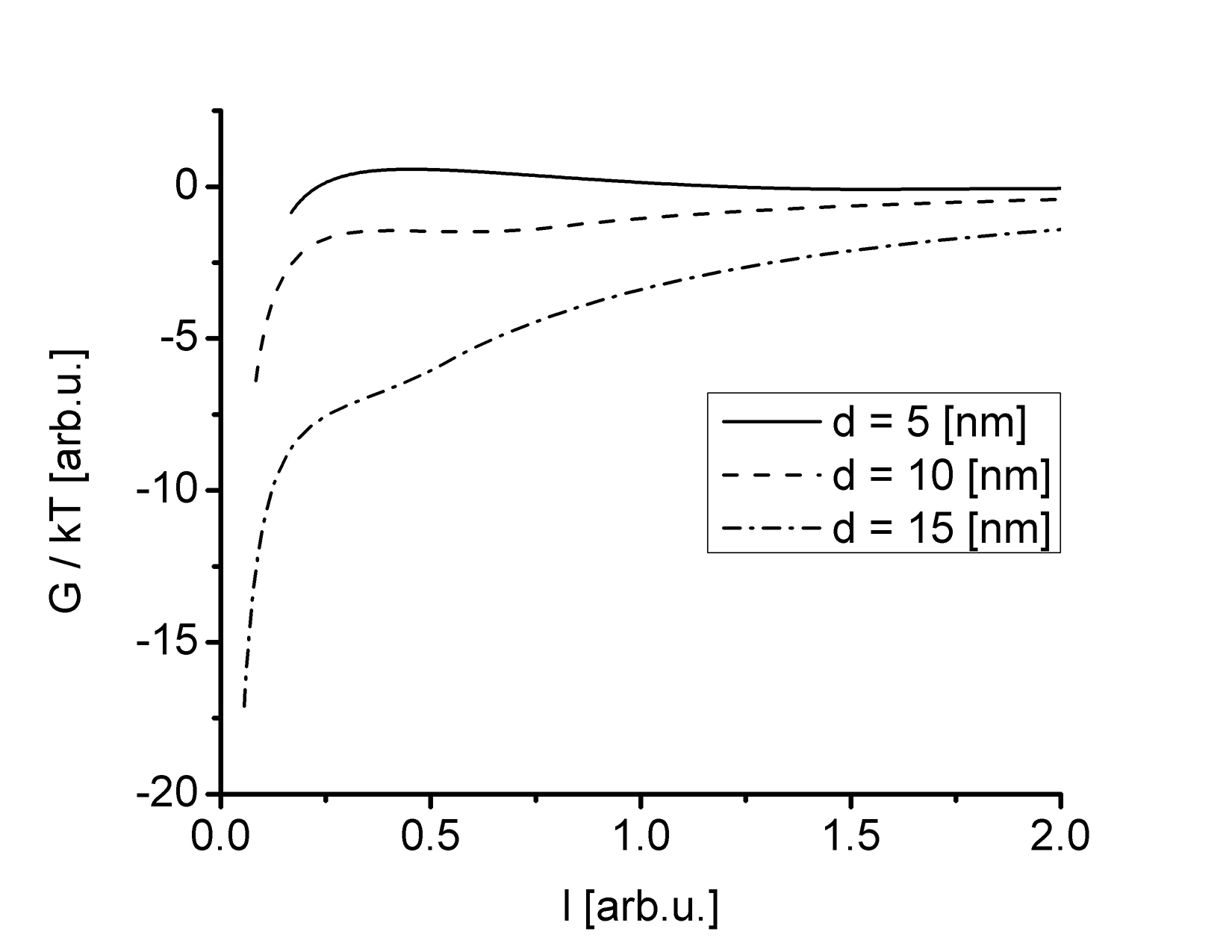}
		\caption{Coating rate by the oleic acid: $k_{\mathrm{c}} = 5\%$}\label{fig:pot_energy_5}
	\end{subfigure}
	\caption{Potential energy dependence on the dimensionless distance between two identical magnetite particles surfaces $l = 2\,s / d$. Different particles diameters $d$ are specified.}\label{fig:pot_energy}
\end{figure}

The saturated surface density of a number of oleic acid molecules in a particle coating is $N_{\mathrm{S}}^{\mathrm{max}} = 2 \cdot 10^{18}\,\mathrm{m}^{-2}$ \cite{Fertman1990}. Depending on a FF preparation recipe variation (an order of mixing/heating of different components \cite{Padalka2004}, etc.), the coating rate $k_{\mathrm{c}} = N_{\mathrm{S}} / N_{\mathrm{S}}^{\mathrm{max}}$ can differ. A classical well stabilized FF usually has $k_{\mathrm{c}} \sim 50\,\%$ which blocks the particles aggregation during durable timeframes (years) due to the free energy barrier $15-25\;kT$ (figure~\ref{fig:pot_energy_50}) \cite{Fertman1990}. The Brownian motion kinetic energy $\sim 1\;kT$ is not enough to overpass the barrier. Only larger particles form aggregates \cite{Zubarev2004b,Tanygin2012e}, which corresponds to the potential well at the $l \approx 0.5 $ (figure~\ref{fig:pot_energy_50}). In the experimental research \cite{Kovalenko2014} the $k_{\mathrm{c}}$ value had been reduced, which leads to the dense phase (drop-like aggregates) emergence (figure~\ref{fig:pot_energy_5}). Same mesoscopic organization control has been reported in the \cite{Lalatonne2005}. The value $k_{\mathrm{c}} = 5\,\%$ has been selected for the present simulation. The calculation of an entropy and the corresponding free energy $F$ by the Eyring's free volume theory \cite{hildebrand1964solubility} based algorithm \cite{Scholten1983} has been made for the system of particles used in the present simulation (figure~\ref{fig:free-energy}). The potential well required for the dense phase emergence is $2-4.5 kT$ or more \cite{Ivanov1996}. An equilibrium state corresponds to the dense phase of such FF. A particles contact leads to an infinite negative potential energy of the van der Waals interaction. The entropic repulsion cannot counteract. However, particles mutual attraction is reversible due to an existence of a minimal distance between their surfaces $s_{\mathrm{min}}\,<\,\delta$ (figure~\ref{fig:pot_energy_5}). The value $s_{\mathrm{min}}\,=\,0.5\,a_0$ nm has been selected as a half of a cell constant which qualitatively reflects restrictions of the Hamaker theory approximated consideration of spherical particles as a continuous body (\ref{eq:van_der_W}) in the case of distances comparable to atomic scale. It corresponds to surface structure peculiarities, a nanoparticle quasicrystalline structure, etc. On the other hand, the value $s_{\mathrm{min}}$ corresponds to an order of magnitude of two oleic acid molecular widths. This additional to the entropic repulsion~(\ref{eq:entropic_rep}) stabilized ``buffer'' role of the surfactant molecule has been discussed in the~\cite{Fertman1990}.

\subsection{Finite-difference scheme}

The finite-difference Euler scheme is based on the original method \cite{Tanygin2012e}. Original method leverages the soft-sphere model for a simulation optimization purpose. The only changes will be described below. Algorithm details of present method can be accessed and contributed in the open-source project ``Ferrofluid Aggregates Nano Simulator'' \cite{Tanygin2014}. The Verlet type of a finite-difference method \cite{Verlet1967} is in need of further method improvement.

A random particle translation and rotation is considered by means of the viscous limit approximation \cite{einstein1905movement,Pottier2014,shliomis1974magnetic} which restricts a time tolerance corresponding to a finite-difference scheme time step:
\begin{equation}
	\Delta t \gg \max_i\left( \tau_{\mathrm{v}}^{\mathrm{T}}, \tau_{\mathrm{v}}^{\mathrm{R}} \right)
\end{equation}
where $\tau_{\mathrm{v}}^{\mathrm{T}} = M_i / \gamma^{\mathrm{T}}_i$ and $\tau_{\mathrm{v}}^{\mathrm{R}} = I_i / \gamma^{\mathrm{R}}_i$. Consequently, an inertial term of the motion equations (\ref{eq:t_motion},~\ref{eq:r_motion}) can be neglected \cite{Pottier2014}. After the integration by the technique \cite{Pottier2014} taking into account~(\ref{eq:rnd_force_1}-\ref{eq:rnd_force_4}) one can obtain particle random motion dispersions $\sigma^{\mathrm{T}}_i \equiv \left\langle \left(\Delta\bm{r}_i\right)^2 \right\rangle$ and $\sigma^{\mathrm{R}}_i \equiv \left\langle \left(\Delta\boldsymbol \varphi_i \right)^2 \right\rangle$ respectively~(cf.~\cite{Wang2002}):
\begin{widetext}
\begin{align}
	&\sigma^{\mathrm{T}}_i\,=\,D^{\mathrm{T}}_i\,\left\{6\,\Delta t\,+3\,\left(M_i\,/\,\gamma_i^{\mathrm{T}}\right)\,\left[-3\,+\,4\,\mathrm{exp}\left(-\gamma^{\mathrm{T}}_i\,\Delta t\,/\,M_i\right)\,-\,\mathrm{exp}\left(-2\,\gamma^{\mathrm{T}}_i\,\Delta t\,/\,M_i\right)\right]\right\} \\
	&\sigma^{\mathrm{R}}_i\,=\,D^{\mathrm{R}}_i \left\{ 6\,\Delta t\,+3\,\left(I_i\,/\,\gamma^{\mathrm{R}}_i\right)\,\left[-3\,+\,4\,\mathrm{exp}\left(-\gamma^{\mathrm{R}}_i\,\Delta t\,/\,I_i\right)\,-\,\mathrm{exp}\left(-2\,\gamma^{\mathrm{R}}_i\,\Delta t\,/\,I_i\right)\right] \right\}
\end{align}
\end{widetext}
where vector $\Delta\boldsymbol\varphi_i$ defines the $i$-th particle rotation (a axis-angle representation; $d\boldsymbol\varphi_i/dt \equiv \boldsymbol \omega_i$); diffusion coefficients are given by the Stokes-Einstein translational and rotational equations: $D^{\mathrm{T}}_i = k T / \gamma_i^{\mathrm{T}}$ and $D^{\mathrm{R}}_i = k T / \gamma_i^{\mathrm{R}}$ respectively. The corresponding normal distribution (Wiener process of Brownian motion) of the $\boldsymbol\xi_i^{\mathrm{T}}$ and $\boldsymbol\xi_i^{\mathrm{R}}$ is calculated by the Mersenne twister algorithm \cite{Matsumoto1998}.

The magnetic moment N\'eel~(\ref{eq:neel_particles}) and Brownian~(\ref{eq:brown_particles}) relaxation types correspond to different calculation algorithms of the magnetic moment motion which were distinguished by a particle diameter criterion. In the case of the N\'eel type and $\Delta t \gg \tau_{\mathrm{N}}$, the magnetic moment was aligned with the total magnetic field direction in the $i$-th particle center $\bm{r}_i$ at each simulation step. In the case of the Brownian type, the rotational motion~(\ref{eq:r_motion}) of the particle with the ``frozen''~(\ref{eq:frozen}) magnetic moment is calculated.

\section{Results and Discussion}
\label{ref_res}

\subsection{Equilibration state calculation}

\begin{figure}[H]
	\centering
	\includegraphics[width=0.5\textwidth]{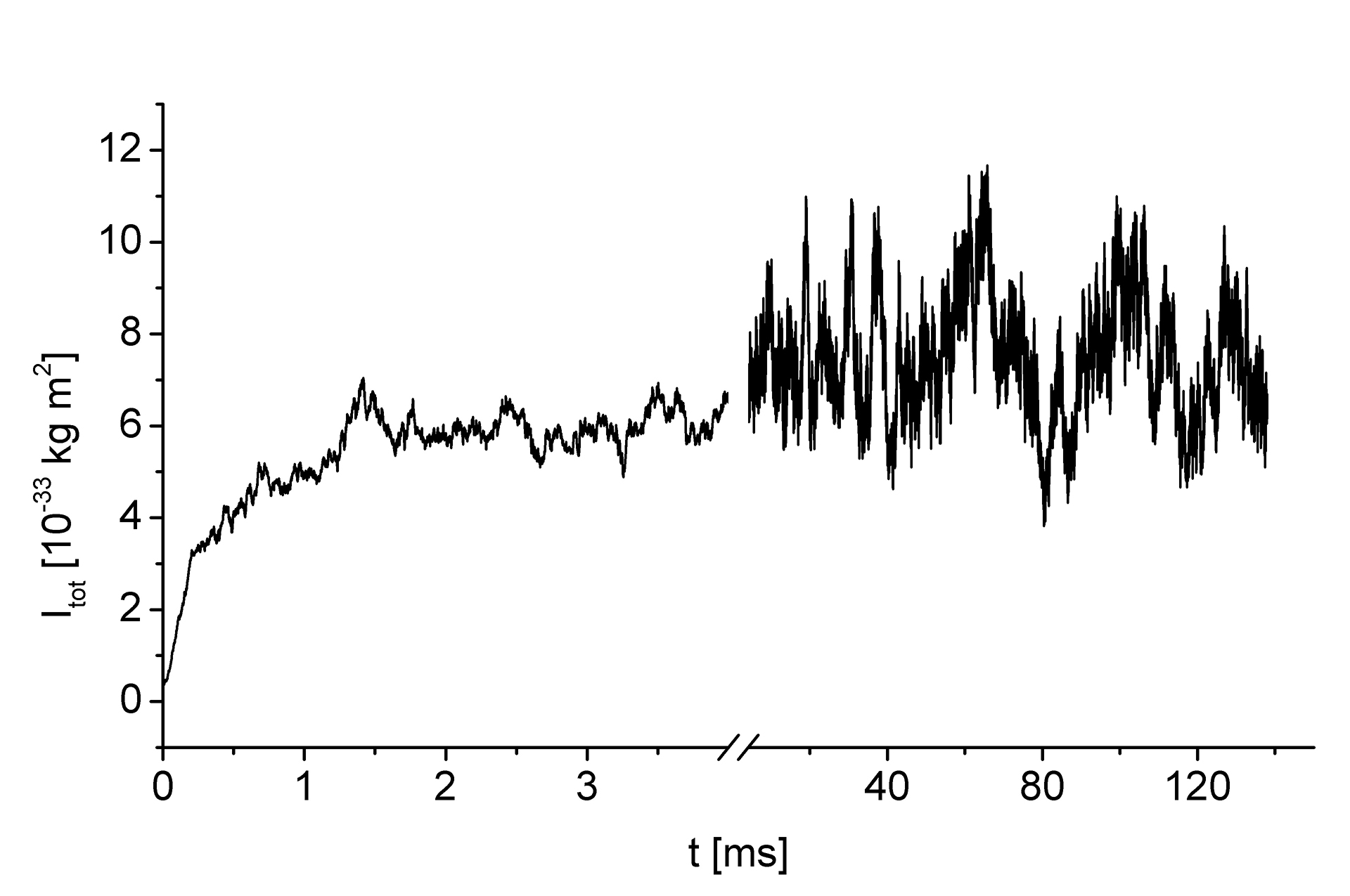}
	\caption{Total moment of inertia time dependence at the conditions: the temperature $T = 25$~$^{\circ}$C and the external field $H_0 = 0$ Oe. The obtained approximated value of the dense phase stabilization period is $t_{\mathrm{s}} \sim 5$ ms.}\label{fig:I(t)}
\end{figure}

An evolution of the initial random close packing structure (figure~\ref{fig:initial-conditions}) to the thermodynamically equilibrium state was calculated. In the case of the dense phase (the nucleus equilibration at the end of the simulation), after a stabilization period $t_{\mathrm{s}}\,=\,t_{\mathrm{s}}\left(H_0,T\right)$, a total moment of inertia of the system $I_{\mathrm{tot}}$ is not changing except its random fluctuations (figure~\ref{fig:I(t)}). This statement has been validated up to $t = 0.2$~s for all obtained dense phases. The resulted structures are shown in the figures~\ref{fig:res-structures-A}~-~\ref{fig:res-structures-B}. In order to distinguish a diluted and dense phase, simpler and similar approach comparing to a conventional correlation functions is used. An one-dimensional chain~\cite{Iskakova2006,Tanygin2012e,Yoon2010,GORENSTEIN2014} is defined in the present simulation as a cluster of particles which satisfies the condition $\left|\boldsymbol\rho_{i\left(i+1\right)}\right| \leq d_i + d_{i+1}$ where $i$-th and $\left(i+1\right)$-th particles are neighbors in the chain. Here, each particle can have only 1 or 2 neighbors. A circle is a particular case of such chain. A set of separated arbitrary length one-dimensional chains is being considered as a diluted phase. If stabilized equilibrium state (see~figure~\ref{fig:I(t)}) corresponds to at least single particle with the number of neighbors larger than 2 then the more complex dense structure is formed. This phase is defined as a dense phase by definition.

\subsection{Phase diagram}

\begin{figure}[H]
	\centering
		\includegraphics[width=0.5\textwidth]{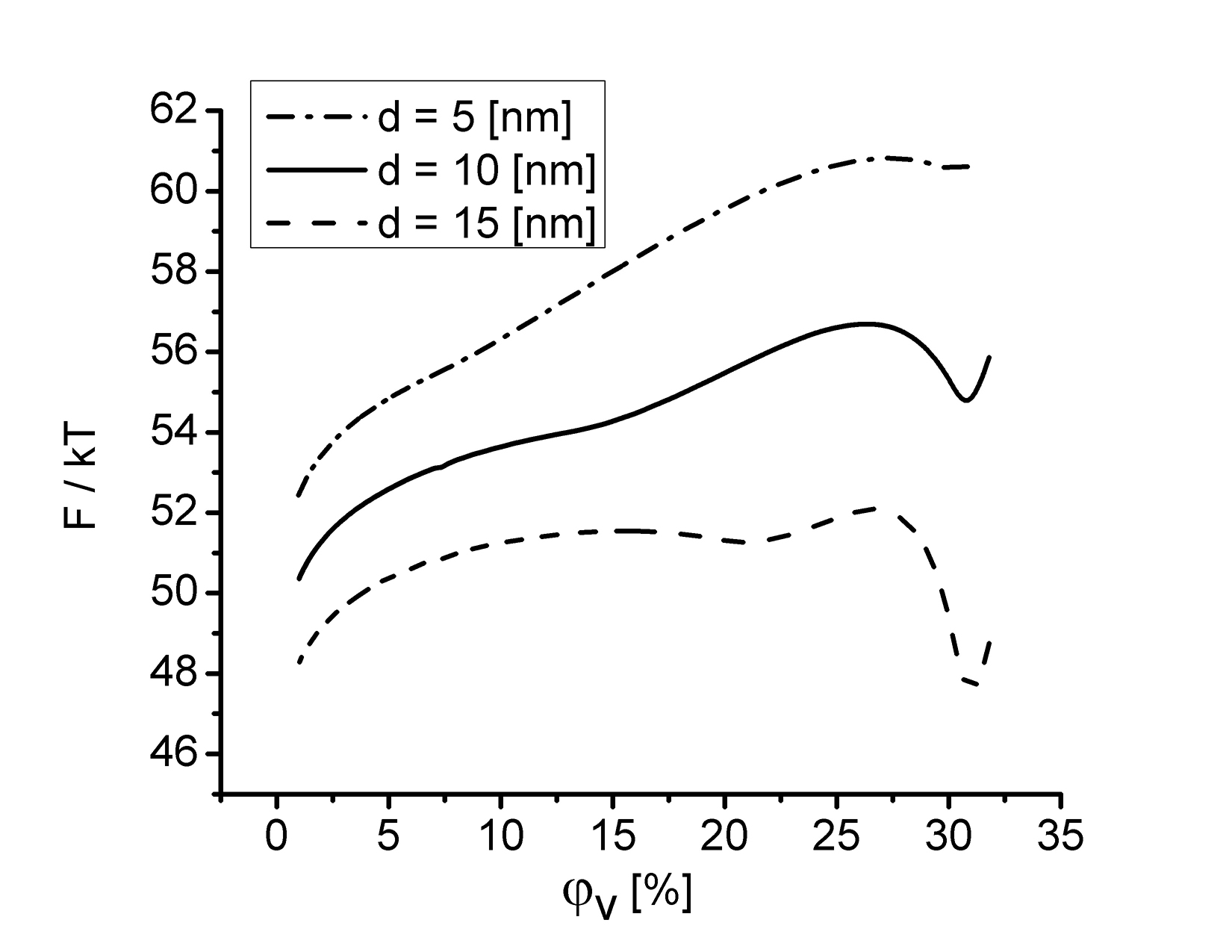}
	\caption{Average free energy of the particle in the aggregate with 100 particles with random positions which fits to the density. The dependence on the volume fraction of the ferrofluid dispersed phase is shown. Coating rate by the oleic acid is $k_{\mathrm{c}} = 5\%$. Different particles $d$ diameters are specified.}\label{fig:free-energy}
	\centering
		\includegraphics[width=0.25\textwidth]{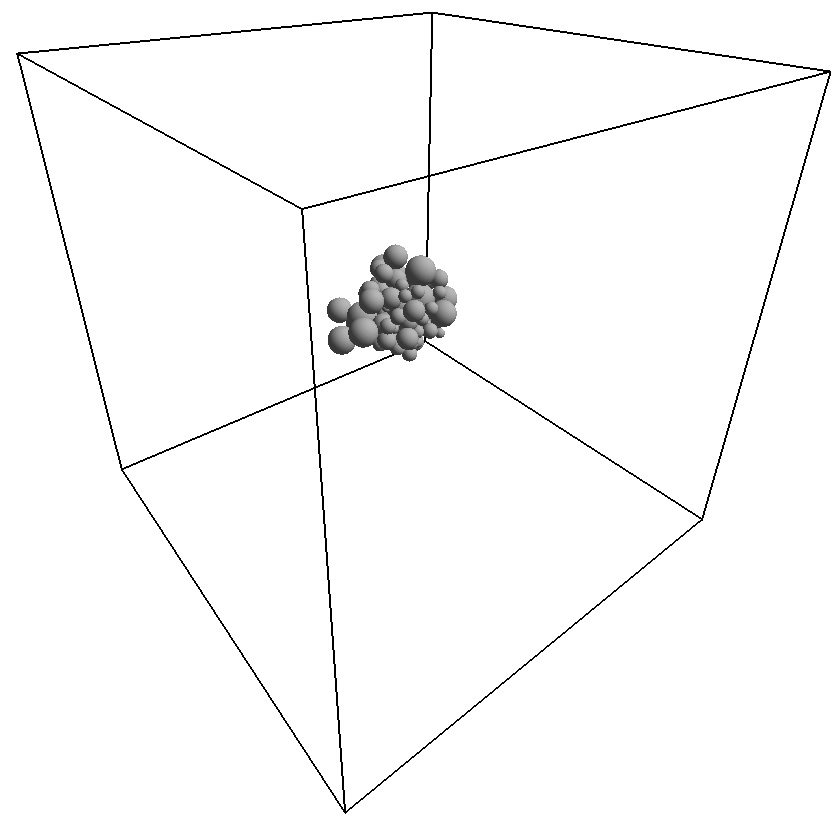}
	\caption{Initial conditions: perspective projection. A space is limited by the cubic vessel with the edge length $L = 0.3$ $\mu$m.}\label{fig:initial-conditions}
\end{figure}

\begin{figure}[H]
	\begin{subfigure}[t]{0.2\textwidth}
		\includegraphics[width=\textwidth]{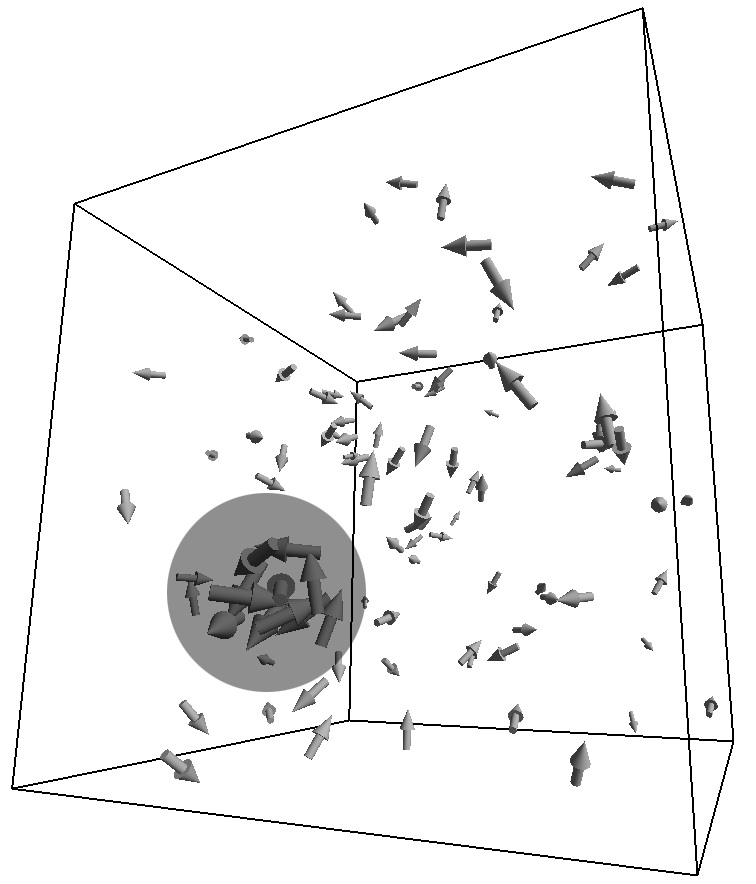} 
		\caption{$H_0 = 0$~Oe and $T = 50$~$^{\circ}$C}\label{fig:H0T50}		
	\end{subfigure}
	\quad
	\begin{subfigure}[t]{0.25\textwidth}
		\includegraphics[width=\textwidth]{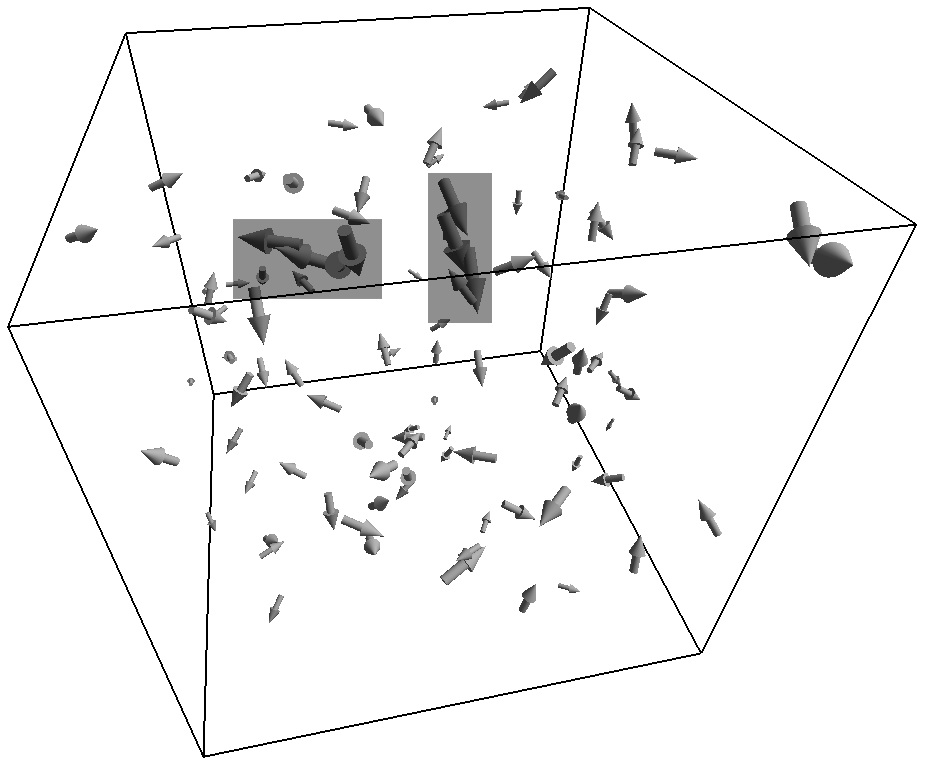} 
		\caption{$H_0 = 0$~Oe and $T = 75$~$^{\circ}$C}\label{fig:H0T75}
	\end{subfigure}
	\quad
		\begin{subfigure}[t]{0.2\textwidth}
		\includegraphics[width=\textwidth]{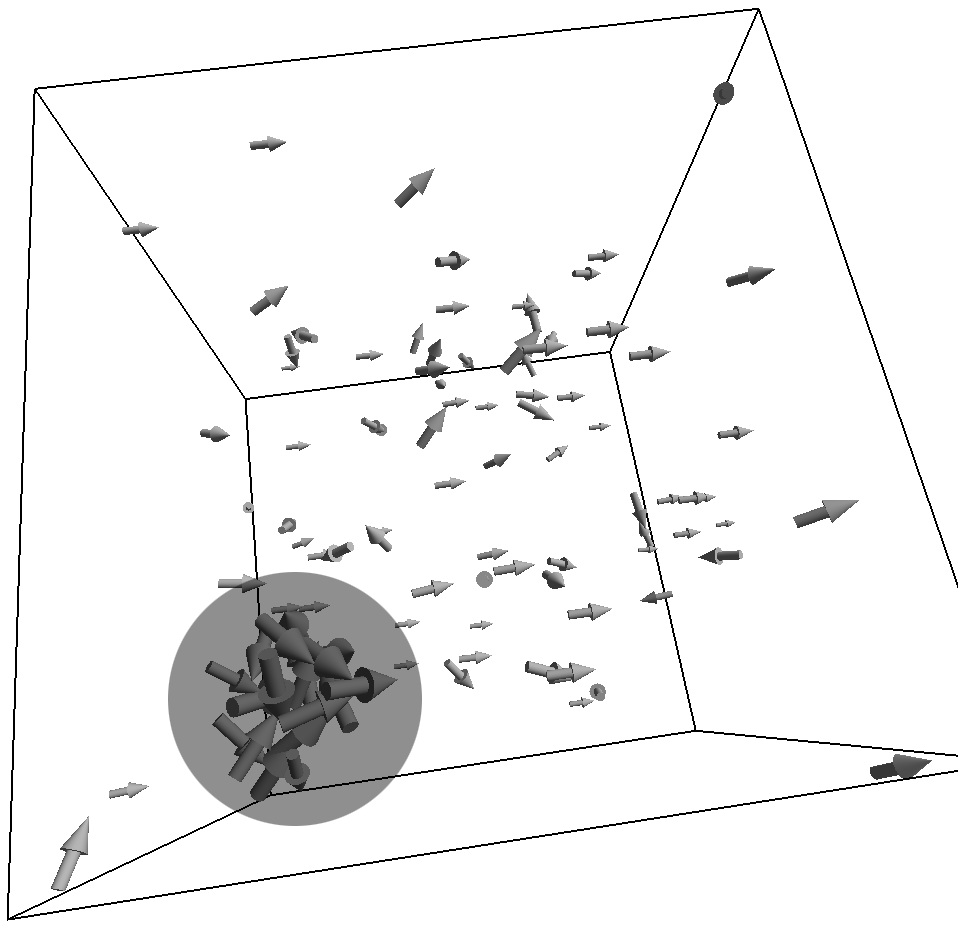} 
		\caption{$H_0 = 200$~Oe and $T = 35$~$^{\circ}$C}\label{fig:H200T35}
	\end{subfigure}
		\quad
	\begin{subfigure}[t]{0.2\textwidth}
		\includegraphics[width=\textwidth]{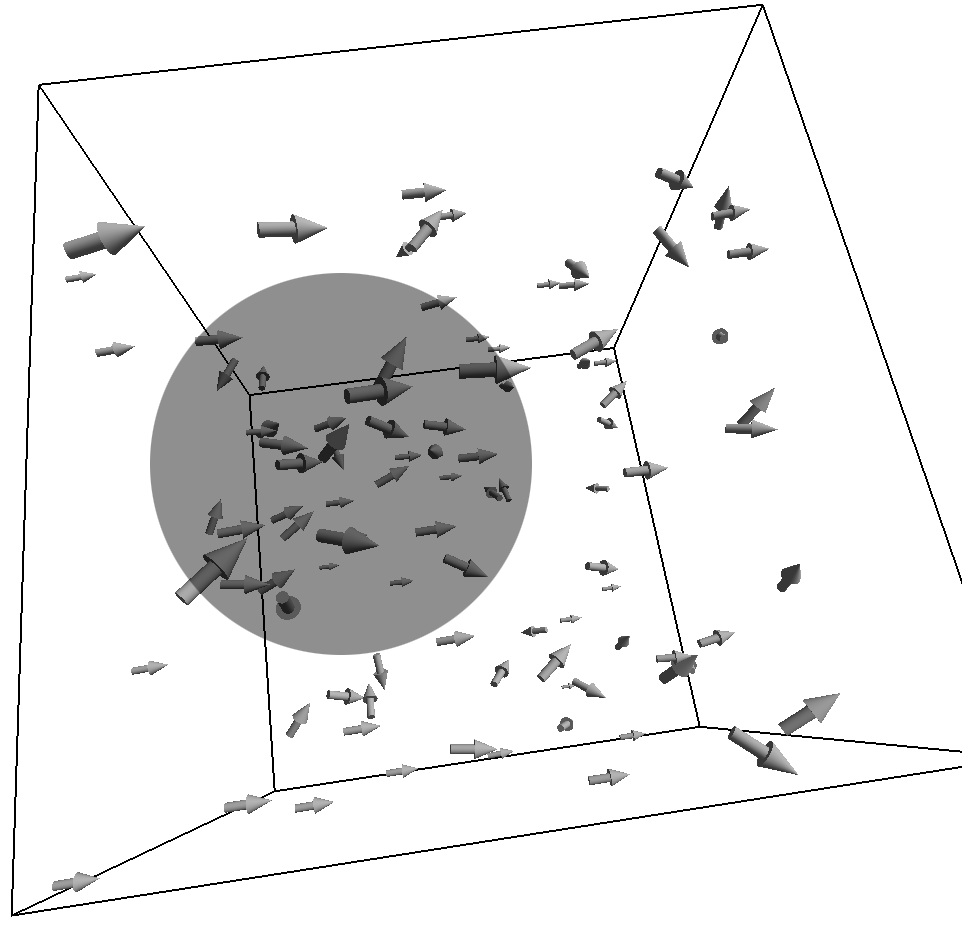} 
		\caption{$H_0 = 200$~Oe and $T = 55$~$^{\circ}$C}\label{fig:H200T55}
	\end{subfigure}
		\quad
	\caption{Resulted thermodynamically equilibrium structures: perspective projections (area with largest particles is highlighted); $L = 0.3$ $\mu$m. A size of volumetric arrows is proportional to a geometrical size of the corresponding nanoparticle for an illustrative purpose only.}\label{fig:res-structures-A}
\end{figure}

\begin{figure}[H]
	\begin{subfigure}[t]{0.2\textwidth}
		\includegraphics[width=\textwidth]{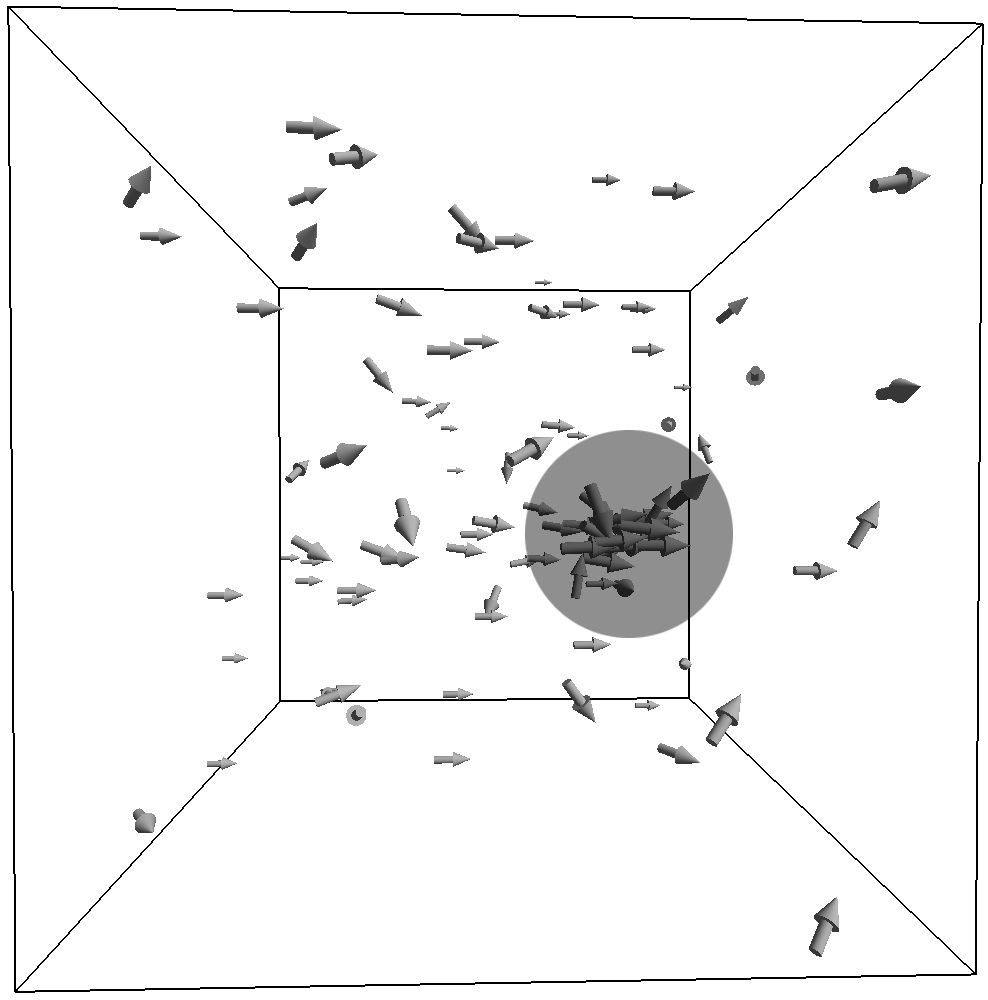} 
		\caption{$H_0 = 400$~Oe and $T = 15$~$^{\circ}$C}\label{fig:H400T15}
	\end{subfigure}
		\quad
	\begin{subfigure}[t]{0.2\textwidth}
		\includegraphics[width=\textwidth]{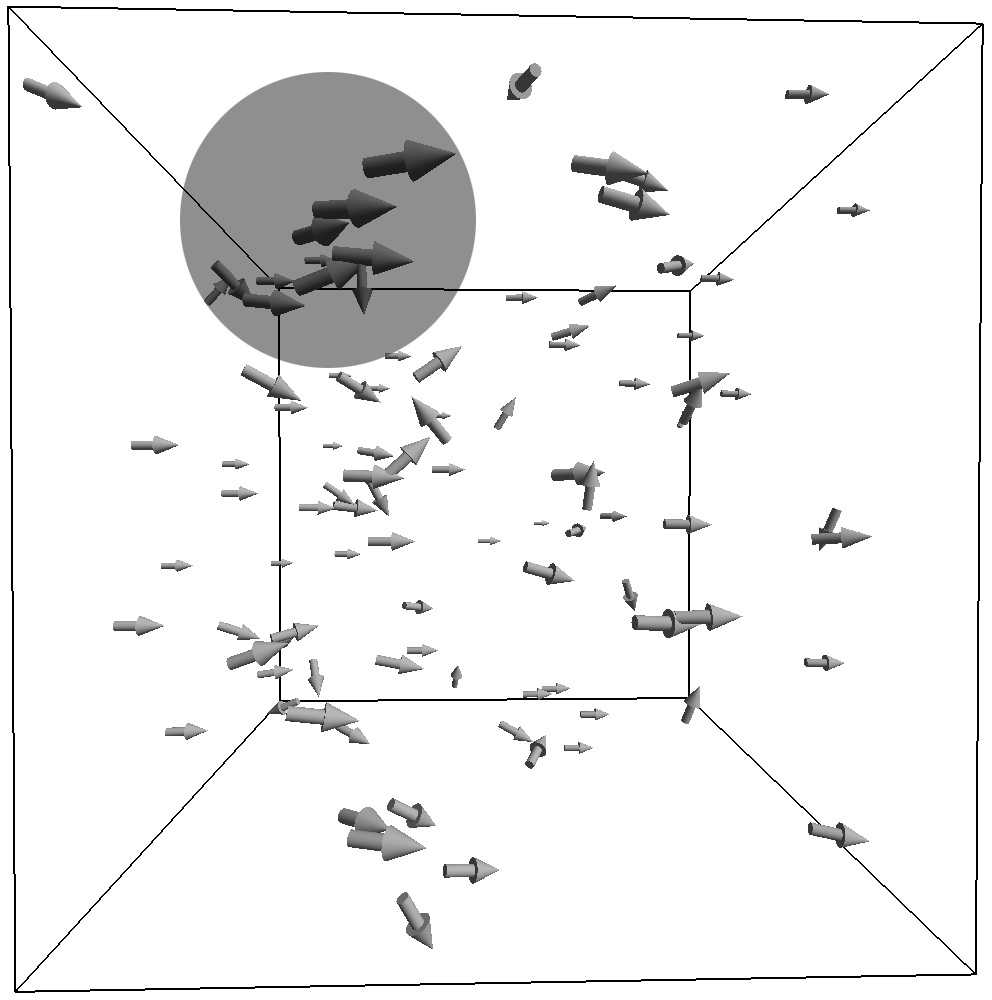} 
		\caption{$H_0 = 400$~Oe and $T = 20$~$^{\circ}$C}\label{fig:H400T20}
	\end{subfigure}
		\quad
	\begin{subfigure}[t]{0.2\textwidth}
		\includegraphics[width=\textwidth]{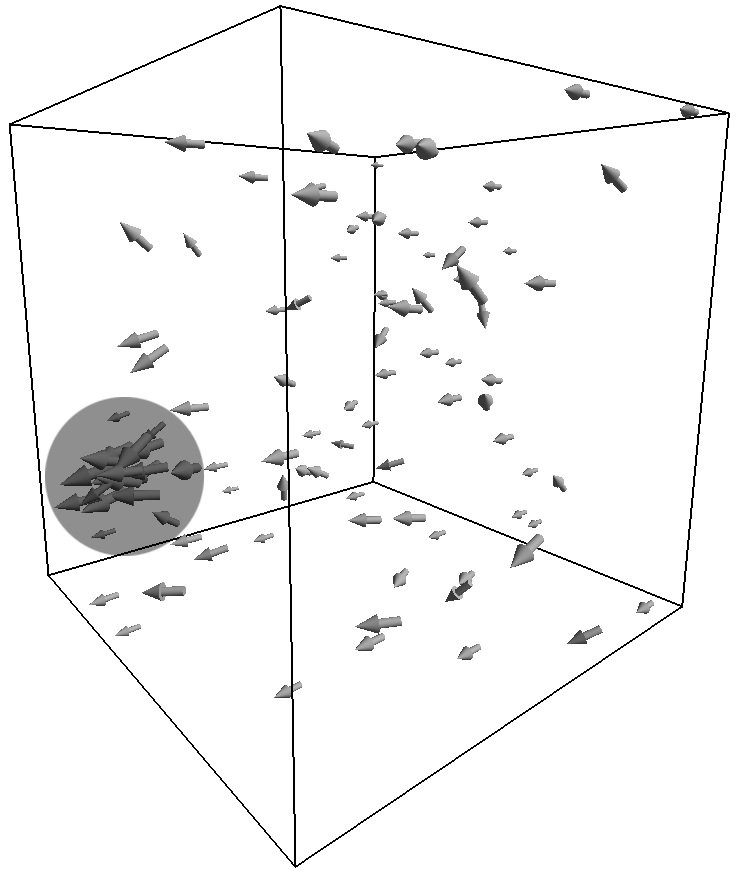} 
		\caption{$H_0 = 800$~Oe and $T = 0$~$^{\circ}$C}\label{fig:H800T0}
	\end{subfigure}
		\quad
	\begin{subfigure}[t]{0.25\textwidth}
		\includegraphics[width=\textwidth]{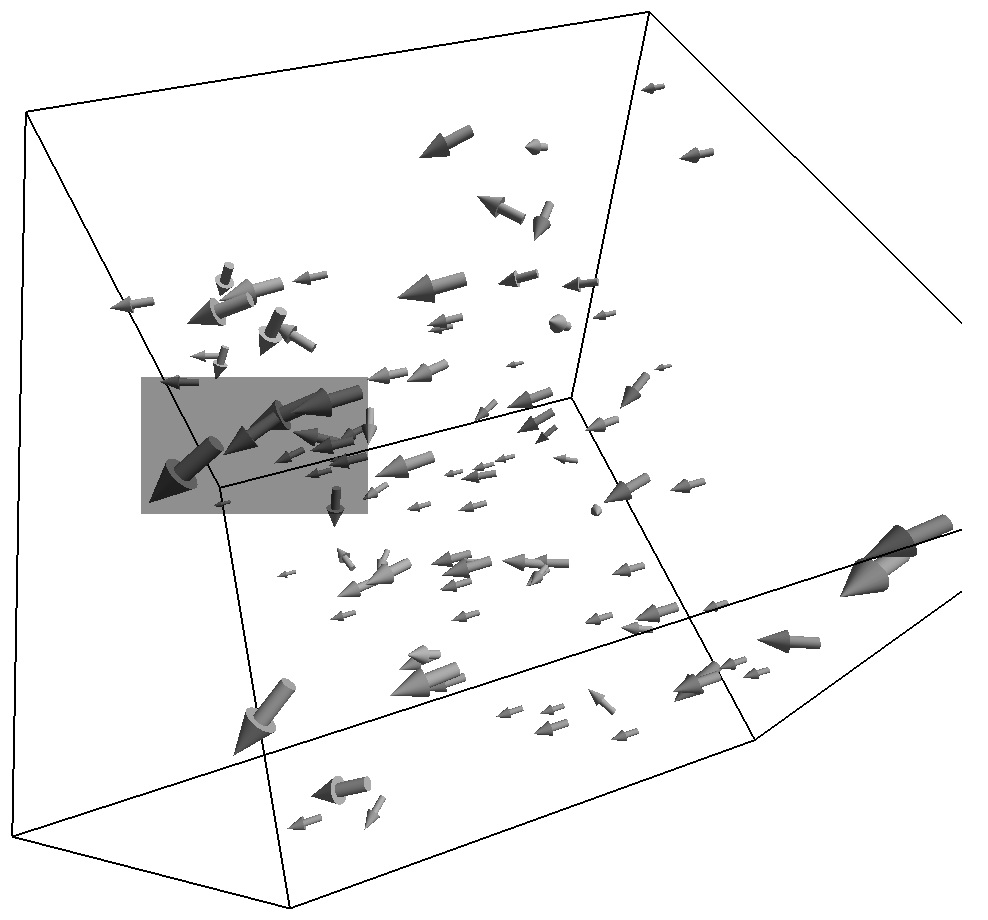} 
		\caption{$H_0 = 800$~Oe and $T = 10$~$^{\circ}$C}\label{fig:H800T10}
	\end{subfigure}
	\caption{Resulted thermodynamically equilibrium structures(continue).}\label{fig:res-structures-B}
\end{figure}

\begin{figure}[H]
	\centering
		\includegraphics[width=0.5\textwidth]{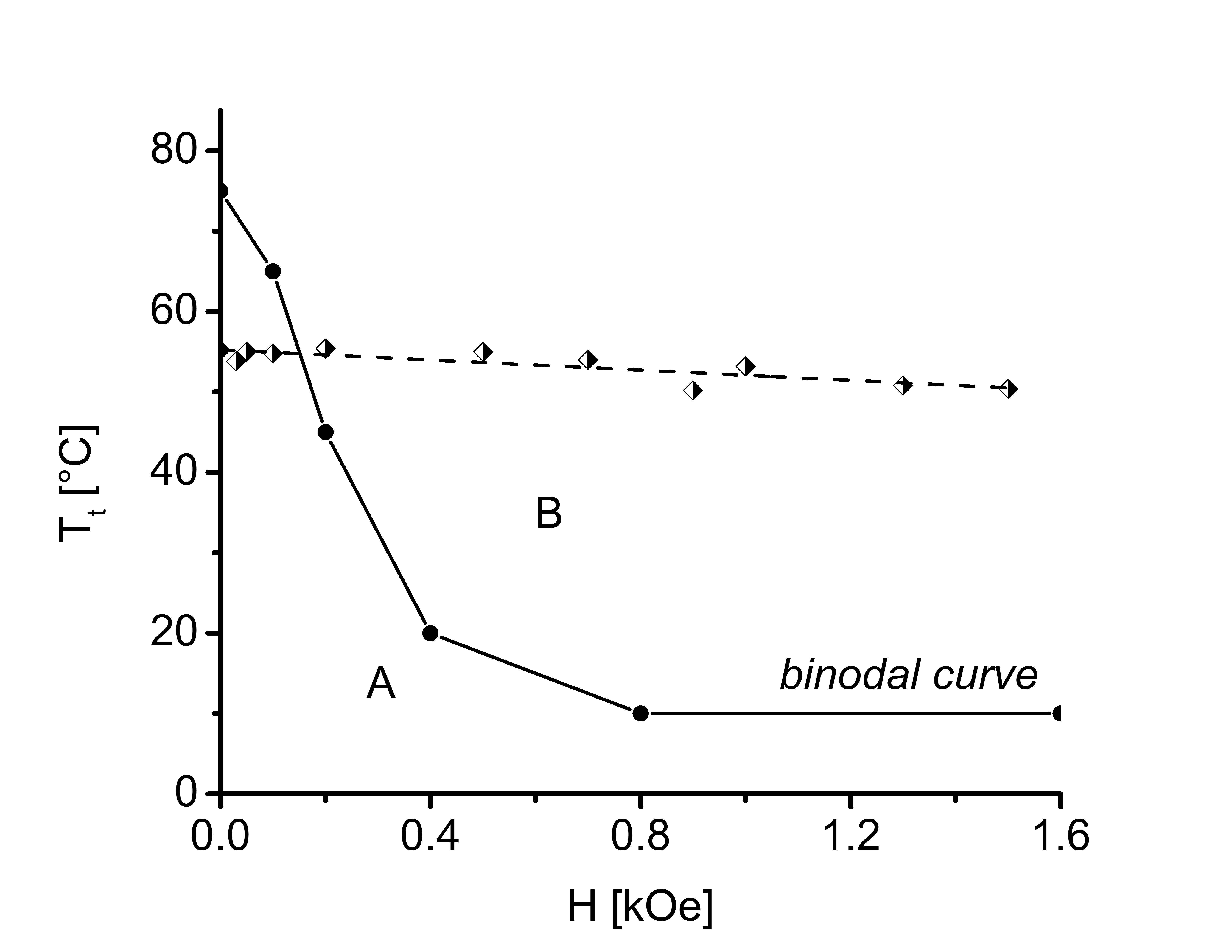}
	\caption{Ferrofluid nucleus phase diagram: a region ``A''\,--\,the aggregated phase and the diluted phase coexistence in the vessel volume; a region ``B''\,--\,the diluted superparamagnetic phase only. The simulation results (circles) form a binodal curve. The experimental results \cite{Kovalenko2014} (diamonds) are given for a comparison.}\label{fig:phase-diagram}
\end{figure}

The phase diagram is obtained based on an analysis of the resulted structures (figures~\ref{fig:res-structures-A}~-~\ref{fig:res-structures-B}) using a bisection method for the definition of the phase transition temperature $T_{\mathrm{t}}$. The binodal curve of the phase diagram (figure~\ref{fig:phase-diagram}) corresponds to the experimental results in terms of the trend and the $T_{\mathrm{t}}$ approximate value in the case of a comparison of temperature values in Kelvins \cite{Kovalenko2013,Kovalenko2014}. This trend has been observed experimentally only starting from the second cycle of the system heating/cooling. The first heating corresponds to the trend weaker than experimental errors and noises \cite{Kovalenko2013,Kovalenko2014}. It can be explained by the assumption that the initial experimental sample state does not correspond to the initial conditions of the simulation i.e. the aggregate with the random close packing \cite{Song2008} of a polydisperse set of particles (figure \ref{fig:initial-conditions}). Indeed, a final equilibrium state of a non-stabilized FF is a set of primary aggregates which consist of large particles only (diameter $d_i^{\mathrm{m}} \sim 15~-~20$ nm) \cite{Tanygin2012e}. Hence, the first heating is required in order to produce the metastable random close packing structure which is kept within several heating/cooling cycles. The large FF aggregate transforms to the primary aggregates state after durable timeframes which require the first heating/cooling cycle again.

The descending binodal curve (figure~\ref{fig:phase-diagram}) contradicts to some theoretical investigations where this dependence is ascending ~\cite{Ivanov1996,Zubarev2004b,Li2007}. As it is stated at the beginning, a contradiction between these theories conclusions and simulation reports is deeper: even a conceptual possibility of an FF l-g transition emergence is being discussed among different research programs \cite{HOLM2005}. Mean-field and statistical models are not justified in the case of a large coupling constant and/or large concentration of FF nanoparticles \cite{Wang2002}. Oppositely, they are justified in the case of a high-temperature and/or a diluted phase FF where a correlation in orientations of magnetic moments is small \cite{Kalikmanov1992}. The latter case corresponds to a modeling of a FF by the classical Lennard-Jones (LJ) fluid and short-range order \cite{Kalikmanov1992,Fisher1964}, which excludes the possibility of an emergence of particles complex long-range ordered structures such as linear chains, rings, tubes, etc. \cite{Yoon2010,Tanygin2012e}. However, it is well established \cite{HOLM2005} that an equilibrium FF phase microstructure consists of a distribution of chains, rings or more complex structures of different lengths. Hence, a real FF dense phase should be modeled by a liquid crystal like microstructure rather than by the classical LJ fluid. However, the LJ fluid is a suitable model for conditions which break the long-range ordered structures through an entropy increase and a predominance of the Brownian motion: a high temperature, a low particles concentration (leading to a structures ``evaporation''), a minor particle coating, etc. In this case a potential energy summand of a FF free energy \cite{Scholten1983} is suppressed by an entropic summand; the free energy local minimum takes place only for large particles (the $\varphi_{\mathrm{V}} < 25~-~30\%$ area in the figure \ref{fig:free-energy}). This is why an l-g transition in the framework of the LJ ferrofluid model is possible only in bi-disperse \cite{Ivanov1996} or polydisperse FF thermodynamic theories.

The LJ like ferrofluid theories correspond to the following idealizations: the continuous isotropic phase gaslike compression model \cite{Li2007}; a considering of a dipole-dipole interaction in scope of the perturbation theory \cite{Ivanov1996}; the mean isotropic potential approximation \cite{Chan1984,Kalikmanov1992,Zubarev2004b}. An external magnetic field \textit{suppresses} the rotational Brownian motion and aligns nanoparticles magnetic moments, which leads to a mean attraction force increase \cite{Chan1984}. Hence, an external magnetic field stimulates the condensation phase transition \cite{Zubarev2004b} leading to the ascending binodal curve. In case of a significant correlation of magnetic moments orientations (non LJ fluid model as discussed above), this conclusion is incorrect because the mean isotropic potential model conditions have been violated: the significantly anisotropic dipole-dipole interaction strongly impacts an equilibrium structure. Such structures with a closed magnetic flux (coils, rings, ring assemblies, tubes, scrolls, etc. \cite{Yoon2010}) already correspond to a \textit{minimal} potential energy and a \textit{suppressed} entropic summand. The applying of an external magnetic field can only increase the potential energy: flux-closed magnetic structures (figure~\ref{fig:H0T50}) transform into the set of parallel linear chains (figures ~\ref{fig:H200T35},~\ref{fig:H400T15},~\ref{fig:H800T0}) where at least marginal particles have a less ``workfunction'' comparing to particles inside the ring or tube structure because they are attached to the single magnetic dipole only. Structures with a closed magnetic flux are bounded by the larger dipole-dipole energy. Such aggregate destructs at a higher temperature.

The present simulation results provide a fine structure \cite{Yoon2010,Tanygin2012e} of the transition from ``linear chains'' to ``dense globes'' (figures~\ref{fig:res-structures-A}~-~\ref{fig:res-structures-B}) through the ring assembly structure \cite{Yoon2010,Tanygin2012e}. Oppositely, the above-mentioned assumptions of the theoretical investigations ~\cite{Ivanov1996,Zubarev2004b,Li2007} suppress this fine structure. The simulation method suggested here does not imply idealizations of the theoretical models.

The last problem which requires discussion is a relation of the opposite results to experimental studies. Both types of binodal dependencies match the respective experimental observations. Hence, their FF parameters were obviously different. The discussed contradiction has been shown only for the FF parameters of the present research \cite{Kovalenko2013,Kovalenko2014}: a small particle coating rate $k_{\mathrm{c}} = 5\%$ by an oleic acid which is a concept similar to the \cite{Lalatonne2005}. This value allows a random compact packing \cite{Song2008} in a polydisperse FF bypassing a strong surfactant repulsion. This is a non-stabilized type of a FF (figure \ref{fig:pot_energy_5}) with a high volume fraction of a dispersed phase: an equilibrium state corresponds to the value $\varphi_{\mathrm{V}} \sim 30\%$. Oppositely, a ``classical'' stabilized FF ($k_{\mathrm{c}} = 50\%$) corresponds to well separated particles (figure \ref{fig:pot_energy_50}). The latter means that a correlation in orientations of magnetic moments is small \cite{Kalikmanov1992} due to a larger average distance between particles and a corresponding weaker dipole-dipole interaction. In this case the aggregates formation can be modeled by an l-g type of a phase transition in the framework of the simple mean field theory \cite{sano1983theory} or the isotropic potential model \cite{Kalikmanov1992,Chan1984}. In the case of a stabilized FF, an ascending binodal curve theoretical dependence ~\cite{Ivanov1996,Zubarev2004b,Li2007} matches very few experimental reports (the reference in the \cite{Ivanov1996}).

This research focuses on the FF aggregate nucleus only. The phase diagram should be verified and generalized to the case of a bulk phase: a larger number of particles and a more durable period of a stabilization. According to the experiment \cite{Kovalenko2013,Kovalenko2014}, this period should have order of magnitude in a range between seconds and minutes time-scales. Taking into account an extensive nature of the current simulation method \cite{Tanygin2014}, its natural further development is a parallel computing capabilities implementation. In our opinion, a most promising method of the parallel computation of the FF molecular dynamics is a Graphics Processing Units based approach \cite{Polyakov2013}. The open-source project dedicated to a further evolution of the method has been started \cite{Tanygin2014}. A comparison of present results with ones based on the Monte Carlo simulation is in need of further investigations~\cite{Satoh1996,Camp2000}.

\section{Conclusions}
\label{ref_con}
\begin{enumerate}
	\item The equilibrium structures of the ferrofluid aggregate nucleus have been determined for different values of temperature and external magnetic field magnitude.
	\item The simulation of the ferrofluid phase diagram, which approximately matches the experiment, is obtained.
	\item The obtained binodal curve has an opposite (descending) trend comparing to some of known theoretical results.
	\item In our opinion, the applied simulation method provides a more justified description of the magnetite ferrofluid with the minor rate of the particle coating by surfactant molecules.
	\item The theoretical concepts which provide the ascending binodal curve dependency trend match other experimental conditions: a diluted ferrofluid, a high particle coating rate, a high temperature, and/or a less particles coupling constant value.
\end{enumerate}

\section*{Acknowledgments}
\label{ref_ack}

We thank Mrs.~Daria~T. for support with the graphical design of the ``FFANS'' web page.

\bibliography{My-Collection}

\pagebreak

\pagebreak

\end{document}